\documentclass[aps,prl,
reprint,
superscriptaddress,
showpacs,
 amsmath,amssymb,
 aps,
]{revtex4-1}

\usepackage{graphicx}
\usepackage{dcolumn}
\usepackage{bm}
\usepackage{color}

\def\bs{\boldsymbol}

\begin{document}


\title{Fast control of topological vortex formation in BEC by counter-diabatic driving}

\author{Shumpei Masuda}
\affiliation{James Franck Institute, The University of Chicago, Chicago, Illinois 60637, USA}

\author{Utkan G\"ung\"ord\"u}
\affiliation{Department of Physics and Astronomy and Nebraska Center for Materials and Nanoscience, University of Nebraska, Lincoln, Nebraska 68588, USA}

\author{Xi Chen}
\affiliation{Department of Physics, Shanghai University, 200444 Shanghai, People's Republic of China}

\author{Tetsuo Ohmi}
\affiliation{Research Center for Quantum Computing and Department of Physics, Kinki University, Higashi-Osaka, 577-8502, Japan}

\author{Mikio Nakahara}
\affiliation{Department of Physics, Shanghai University, 200444 Shanghai, People's Republic of China}
\affiliation{Research Center for Quantum Computing and Department of Physics, Kinki University, Higashi-Osaka, 577-8502, Japan}

\date{\today}

\begin{abstract}
Topological vortex formation has been known as the simplest method for vortex formation in BEC of alkali atoms. This scheme requires inversion
of the bias magnetic field along the axis of the condensate, which leads to atom loss when the bias field crosses zero. In this Letter, we propose a scheme with which
the atom loss is greatly suppressed by adding counter-diabatic magnetic field. A naive counter-diabatic field violates the Maxwell equations and we need to introduce
an approximation to make it physically feasible. The resulting field requires an extra currents, which is experimentally challenging. Finally we solve this problem
by applying a gauge transformation so that the counter-diabatic field is generated by controlling the original trap field with the additional control of the
bias field.

\end{abstract}

\maketitle


Demonstration of quantized vortices in a superfluid is a manifestation of the
nonvanishing order parameter. Topological structure of vortices often
reflects the manifold of the order parameter space.
It is, therefore, natural to seek for the method
to experimentally demonstrate formation of vortices once BEC of alkali metal
atoms has been realized. Topological method of vortex formation is the simplest one
compared with other proposals, where the spinor structure of the order parameter
is fully utilized \cite{top1,top2,top3,top4,Pietila2007,toprev,Kuopanportti2013}.
Although the vortex \cite{toprev} and monopole \cite{Ray2014,Ray2015} in BEC have been observed experimentally, establishment of methods to create such interesting quantum states more accurately in time much shorter than coherence time is important for further experimental study and  applications.

A condensate in the weak field seeking state (WFSS)
 is trapped in a quadrupole magnetic field
in the $xy$-plane with a uniform bias field along the $z$-axis first and,
subsequently,
the bias field is reversed from $B_z$ to $-B_z$ adiabatically. Then a vortex
with the winding number $2F$ is formed, were $F$ is the quantum number of
the hyperfine spin of the condensate.
This theoretical proposal has been demonstrated experimentally by several groups
\cite{exp1,exp15,exp2,exp25,exp3,exp4}.
A part of the condensate is lost while vortex formation takes place. This is due to
the fact that the gap between the WFSS and the neutral state (NS) and the strong
field seeking state (SFSS) is gradually closed near the center of the condensate during the vortex formation and the
atoms in the trapping WFSS make transition to the NS and the SFSS through the Majorana
flops. It is the purpose of this Letter to show that it is possible to suppress the
Majorana flops so that more atoms
are left in the trap when a vortex is formed.
It is certainly desirable to have more atoms left in the trap since
higher atomic density implies (i) easier measurement of physical
quantities (ii) fighting against atom loss due to hyperfine spin changing collision
and (iii) smaller vortex core size, which is advantageous for
forming a vortex in an arbitrary position in the condensate \cite{yk}.
Motivated by the concept of shortcuts to adiabaticity \cite{review}, we use the counter-diabatic formalism (or quantum transitionless algorithm)
developed by Demirplak and Rice \cite{cd1} and formulated by Berry \cite{cd2,cd3} and demonstrated experimentally in \cite{Bason}.
The supplementary counter-diabatic field (CDF) thus obtained,
is, however, unphysical in that it violates the Maxwell equation div$~\boldsymbol{B}=0$
and some approximation must be introduced so that the counter-diabatic control is
physically feasible. The approximated CDF turns out to be a quadrupole
field, which can be implemented with an ordinary Ioffe bars. However, the CDF is generated by another set of Ioffe bars, which is obtained by
rotating the original confining Ioffe bars by $\pi/4$.
This means that two sets of Ioffe bars are required to physically realize this nonadiabatic control, which makes its experimental
demonstration rather demanding. To overcome this difficulty, we apply the gauge
transformation \cite{sara} to make the combined magnetic field (the original confining field
and the CDF superposed)
parallel to the original confining quadrupole field. The modest price one
has to pay by this gauge transformation is to modify the time-dependence of the
bias field $B_z(t)$ from a linear behavior.
Experimental demonstration is made much easier with this modification.

We first regard each atom with hyperspin $\boldsymbol{F}$ as an independent
quantum system fixed at $\bs{r}$ and consider nonadiabatic control thereof. Let
$\boldsymbol{B}(\bs{r}, t)=(\bs{B}_{\perp}(\bs{r}), B_z(t))$ be the local magnetic field with
the quadrupole field $\bs{B}_{\perp}(\bs{r})=B'_{\perp} (x, - y)$ and
the uniform bias field $B_z(t) = B_z(1-2t/T)$ for $0\le t\le T$, $B_z(t)=-B_z(0)$ for $t> T$ where $T$ is the inversion time
of $B_z(t)$. An atom at $(\bs{r}, t)$ interacts with the magnetic
field through the Hamiltonian
\begin{equation}\label{eq:ham0}
H_B(\bs{r}, t) = \gamma \bs{B}(\bs{r}, t)\cdot \bs{F} ,
\end{equation}
where $\gamma=\mu_B g_F$, $\mu_B$ is the Bohr magneton, $g_F$ is the $g$-factor for
the hyperfine spin $F$ and $\bs{F}=(F_x, F_y, F_z)$ is $(2F+1)$-dimensional
irreducbile representation of the Lie algebra $\mathfrak{su}(2)$.
We take F = 1 in the following for definiteness.
For a given $\bs{B}(\bs{r},t)$, the Hamiltonian (\ref{eq:ham0}) has three normalized eigenstates
\begin{eqnarray}
|\mbox{WFSS} \rangle &=& \frac{1}{2B}
\begin{pmatrix}
 (B - B_z)e^{2i\phi} \\
- \sqrt{2}B_\bot e^{i\phi}\\
 B + B_z
\end{pmatrix},
\nonumber\\
|\mbox{NS} \rangle &=&
\frac{1}{\sqrt{2}B} \begin{pmatrix}
-B_\bot e^{2i\phi} \\
\sqrt{2}B_z e^{i\phi} \\
B_\bot
\end{pmatrix}, \label{NS}
\\
|\mbox{SFSS} \rangle &=& \frac{1}{2B}\begin{pmatrix}
 (B + B_z)e^{2i\phi} \\
\sqrt{2}B_\bot e^{i\phi}\\
 B - B_z
\end{pmatrix},
\nonumber
\end{eqnarray}
where $B = \sqrt{B_{\perp}^2(r)+B_z^2(t)}$ and $\phi$ is the azimuthal angle.
The corresponding eigenvalues
are $|\gamma| B$, $0$, and $-|\gamma| B$, respectively,
where $\gamma = -\mu_B/2$. The WFSS with
the  highest energy $|\gamma| B$
is the only state that can be magnetically trapped. We start with a pure WFSS at $t=0$ and
want to keep as many atoms in the WFSS as possible at $t=T$, when the vortex formation is
complete, in spite of vanishing gaps among the three states at the origin ($r=0$)
at $t = T/2$. This is realized by adding a CDF on the atom.

Here we briefly summarize the counter-diabatic approach to nonadiabatic quantum control
and apply it to our hyperfine spin control. Let $H_0(t)$ be a time-dependent Hamiltonian
and $|n(t) \rangle$ be an instantaneous eigenvector with the eigenvalue $E_n$ such that $H_0(t)|n(t) \rangle
= E_n(t)|n(t) \rangle$. In the adiabatic approximation, the solution to the time
dependent Schr\"odinger equation is $|\psi_n(t) \rangle = e^{-i \gamma_n(t)}|n(t) \rangle$,
where
\begin{equation}
\gamma_n(t) = \frac{1}{\hbar}\int_0^t dt' E_n(t') -i \int_0^t dt' \langle n(t')|\partial_{t'}n(t') \rangle.
\label{eq_gamma}
\end{equation}
Let $U(t) = \sum_n |\psi_n(t)\rangle \langle n(0)|$ be the time-evolution
operator such that $U(t):|n(0) \rangle \mapsto |\psi_n(t) \rangle$. The operator
$U(t)$ defines a Hamiltonian $H(t) = i\hbar(\partial_t U(t))U^{\dagger}(t)$
for an arbitrary
time evolution, not necessarily adiabatic, namely $|\psi_n(t)\rangle$ is the {\it exact}
solution of
\begin{equation}
i \hbar\partial_t|\psi_n(t) \rangle = H(t)|\psi_n(t) \rangle,
\end{equation}
where $|\psi_n(0) \rangle =|n(0) \rangle$ is satisfied by definition.
If we write $H(t) =H_0(t) + H_{\rm CD}(t)$, the counter-diabatic Hamiltonian
$H_{\rm CD}(t)$ is written as
\begin{eqnarray}
H_{\rm CD}(t) &=& i \hbar \sum_n |\partial_t n(t)\rangle \langle n(t)|,
\label{HCD}
\end{eqnarray}
when the ``parallel" condition, $\langle n(t)|\partial_t n(t) \rangle =0 $, is satisfied. 

Let us apply the counter-diabatic scheme to
our hyperfine spin system
by taking $H_0$ and $|n\rangle$ as $H_0(\bs{r},t)=H_B(\bs{r},t)$ and
the instantaneous eigenstates in Eqs. (\ref{NS}).
Note that the coordinate $\bs{r}$ here is
just a parameter specifying the position of the atom in the condensate, and the geometric phase in Eq. (\ref{eq_gamma}) vanishes.
By substituting Eq. (\ref{NS}) into (\ref{HCD}), the counter-diabatic Hamiltonian is obtained as
\begin{equation}
H_{\rm CD} = \gamma \bs{B}_{\rm CD} \cdot  \bs{F},
\label{hanacd1}
\end{equation}
where
\begin{eqnarray}
\bs{B}_{\rm CD}(\bs{r}, t) &=&
\frac{-2}{\gamma T} \frac{B_z(0) B'_{\perp}}{B^2(r, t)}(y,x,0) \label{eq:bcd}
\end{eqnarray}
is the CDF.
This magnetic field looks as if it could be generated by four Ioffe bars that are obtained by
rotating the original Ioffe bars by an angle $\pi/4$ around the $z$-axis.
An important remark is in order here. When the counter-diabatic scheme is applied
to a quantum system, it produces a counter-diabatic potential (the magnetic
field in our case) that prevents the quantum system from escaping from an
adiabatic time evolution. There is no guarantee, however, that the counter-diabatic
potential will be physically feasible. In fact,
$\bs{B}_{\rm CD}$ in Eq. (\ref{eq:bcd}) does not satisfy ${\mathrm{div}}\bs{B}_{\rm CD}=0$. This is not unexpected
since we did not consider the Maxwell equation while we obtained $\bs{B}_{\rm CD}$.
The way out of this problem is to fix the coordinate $r$ to $r_0$ in $B^2(r, t)$
in the denominator of $\bs{B}(r, t)$ so that $\bs{B}_{\rm CD} (r,t)
\propto (y,x,0)$ and ${\mathrm{div}}\bs{B}_{\rm CD}$ vanishes.
In our calculation below,
we show that the CDF increases the number of atoms forming the vortex in a wide range of the inversion time $T$ by using
several values of $r_0$.
The performance may be further improved if we optimize the number of
atoms left in the trap by varying $r_0$, which will be worked out in our future
publication. We call the combined magnetic field as $\tilde{\bs{B}}
= \bs{B}+\bs{B}_{\rm CD}$.

Now we solve the Gross-Pitaevskii
equation (GPE) with the designed magnetic field $\tilde{\bs{B}}$;
 \begin{eqnarray}  
 i \hbar\partial_t  \Psi_m(r, t) &=& \Big\{h_{mn} + g_n \delta_{mn} \sum_p |\Psi_p|^2
\nonumber\\
& &  + g_s \sum_{\alpha} \sum_{l,p} \left( \Psi_l (F_{\alpha})_{lp}\Psi_p \right)
 (F_{\alpha})_{mn}\Big\} \Psi_n,\nonumber\\
 & & \label{eq:GP1}
 \end{eqnarray}
 where $l, m,n,p \in \{-1,0,1\}$, $\alpha \in \{x,y,z\}$,
 $$
 h_{mn} = \ \Big{(}-\frac{\hbar^2\nabla^2}{2M} - \mu\Big{)} \delta_{mn} + \mathcal{B}_{mn},~  \quad
 {\mathcal{B}} =\gamma \tilde{\bs{B}} \cdot \bs{F},
 $$
with the mass of the atom $M$, and the chemical potential $\mu$ (the eigenvalue of the GPE at $t=0$).
 To begin with, we need to find the initial condition to solve the GPE (\ref{eq:GP1}).
 Assuming that the initial state is in the WFSS  $|\Psi(0) \rangle= f(\bs{r})|{\rm WFSS}(0)\rangle$,
 the condensate wave function is obtained
 by solving the stationary GPE
 \begin{eqnarray}\label{eq:GP0}
& &-\frac{\hbar^2}{2M} \Big[ \frac{1}{r} \frac{d}{dr}\left(r\frac{d}{dr}\right) -\frac{\beta'^2}{2}
 -\frac{1}{4r^2} (7-8 \cos \beta + \cos 2 \beta)\Big] f(r) \nonumber\\
 & &\qquad  + \gamma \tilde{B}(\bs{r},0) f(r)  + g f^3(r) = \mu f(r),
 \end{eqnarray}
where $g=g_n+g_s$,
$\tilde{B}(\bs{r}, t) = \{[\bs{B}_{\perp}(\bs{r})+\bs{B}_{\rm CD}(\bs{r},t)]^2+ B^2_z(t) \}^{1/2}$,
and the WFSS is parametrized as
 $$
 |{\rm WFSS} \rangle = \left(\begin{matrix}
 \cos \beta \cos \phi  - i \sin \phi\\
 -\cos \beta \sin \phi - i \cos \phi\\
 -\sin \beta
 \end{matrix} \right),
$$
with
$$
\beta = \tan^{-1} \left[\frac{|\bs{B}_{\perp}(\bs{r})+\bs{B}_{\rm CD}(\bs{r},t)|}{B_z(t)} \right].
$$
Here we note that $f(\bs{r})$ and $|\bs{B}_{\perp}(\bs{r})+\bs{B}_{\rm CD}(\bs{r},0)|$ are
in fact the functions of $r$ only. It turns out that $\bs{B}_{\rm CD}$ is negligibly small
at $t=0$ and it can be ignored safely in solving Eq.~(\ref{eq:GP0}).
The time-dependent GPE is solved numerically and
we summarize the results below. Throughout our calculation, we have taken the parameters
of $^{23}$Na atoms \cite{top3}, $g_{n}=0.0378 a_{\rm HO}^3\hbar\omega$, $g_s=0$
and $B_z(0) =1$~G, $B'_{\perp} = 300$~G/cm for which
the harmonic oscillator length is found to be $a_{\rm HO} \sim 9.14 \times 10^{-1}~\mu$m and
$\hbar \omega \sim 3.49 \times 10^{-24}$~erg. Lengths and energies are scaled by
these parameters. In addition, we find $\omega_L(r=0,t=0) \sim 4.40 \times 10^6$~rad/s.
The time scale $\tau = 2\pi/\omega_L(0,0) \sim 1.43~\mu$s is a reasonable measure of adiabaticity.
The chemical potential is found to be $\mu - \hbar \omega_L(0,0) \sim 3.66$ in dimensionless units.

Figure 1 shows the time dependence of the ratio $|\bs{B}_{\rm CD}|/|\bs{B}_{\perp}|$ for $\log_{10}(T/\tau)=1.4$ and $r_0/a_{\rm HO} = 2$, which is independent of $r$ and depends only on $t$. The inset is the
profile of the condensate at $t=0$, which is obtained by solving
Eq.~(\ref{eq:GP0}) numerically.
The units of time, length and $f$ are $\tau$, $a_{\rm HO}$ and $a_{\rm HO}^{-3/2}$,
respectively.
\begin{figure}
\begin{center}
\includegraphics[width=6cm]{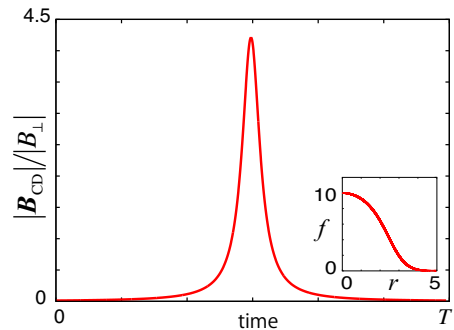}
\end{center}
\caption{(Color online) Time dependence of the ratio  $|\bs{B}_{\rm CD}|/|\bs{B}_{\perp}|$,
which is independent of $r$ and depends only on $t$. Inset shows
the bound state wave function obtained by solving Eq.~(\ref{eq:GP0}) for $^{23}$Na.}
\end{figure}
Now the GP equation (\ref{eq:GP1}) is solved to
find interesting quantities. Let us show first how much atoms are left in the trap
after the vortex formation takes place.
To mimic the trap loss, we multiply the order parameter by
$h(r) = 0.5[1-\tanh((r-r_1)/\lambda)]$ with $r_1=30a_{\rm HO}$ and $\lambda=2a_{\rm HO}$ at each time step of the numerical simulation \cite{top3}.
In Fig. 2 we show the snapshots of the amplitude of the order parameter $\sum_n|\Psi_n|^2$ for $0\le t \le 100T$ with the CDF with the same parameter set as Fig. 1.
The condensate in the trapping region at $t=100T$ is mostly in WFSS, while SFSS and NS are gradually removed from the trap in the domain $r>r_1$ due to $h(r)$.
The zero of the order parameter at $r=0$, $t=100T$ is a manifestation of
the vortex of the WFSS.
Fig. 3 shows the fraction of atoms $N(t)/N(0)$ left
in the trap at $t \gg T$. Since we replaced an unphysical $\bs{B}_{\rm CD}$ with a physical
$\bs{B}_{\rm CD}$ satisfying ${\mathrm{div}}~\bs{B}_{\rm CD}=0$, the result depends on the
parameter $r_0$ in the denominator of Eq.~(\ref{eq:bcd}).
\begin{figure}
\begin{center}
\includegraphics[width=6cm]{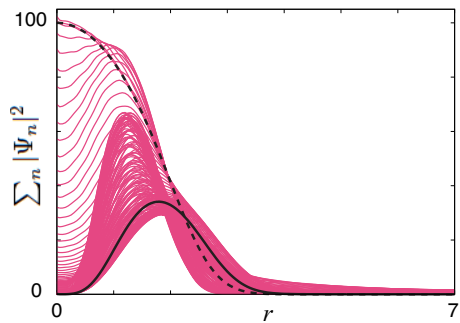}
\end{center}
\caption{(Color online) Snapshots of the amplitude of the order parameter
for $0\le t \le 100T$ with the same parameter set as that in Fig. 1.
Time intervals between snapshots are equal with each other.
The dashed and thick solid curves correspond to $t=0$ and $t=100T$, respectively. The unit of $\sum_n|\Psi_n|^2$ is $a_{\rm HO}^{-3}$.}
\end{figure}
\begin{figure}
\begin{center}
\includegraphics[width=7cm]{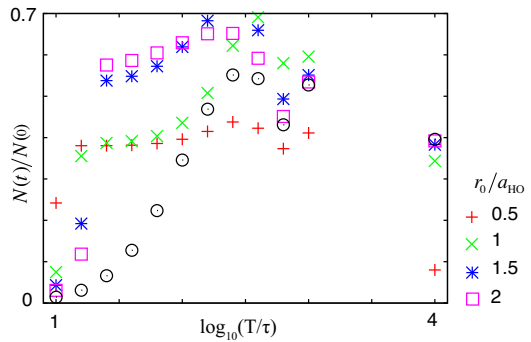}
\end{center}
\caption{(Color online) Fraction $N(t)/N(0)$ of atoms left in the trap long time after a vortex is
formed. Since $t \gg T$, the condensate is in the pure WFSS. $\circ$ shows
the fraction without $\bs{B}_{\rm CD}$ while other symbols show the fractions with
$\bs{B}_{\rm CD}$ for different $r_0$.}
\end{figure}
Observe the prominent improvement in the ratio $N(t)/N(0)$ for small $T$.
This is more clearly seen by plotting $N(t)$ in the control with the CDF normalized
by $N(t)$ in the control without the CDF. Figure 4 shows the ratios with the same
$r_0$ as Fig. 3. The ratio is almost 20 for $T=10 \tau$ and $r_0=0.5 a_{\rm HO}$.
\begin{figure}
\begin{center}
\includegraphics[width=6cm]{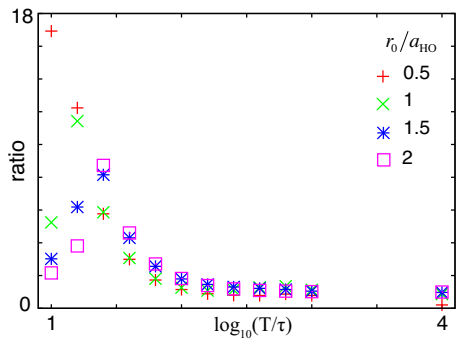}
\end{center}
\caption{(Color online) Number of atoms $N(t)$ for $t\gg T$ in the control with the CDF normalized by $N(t)$ in the control without the CDF. The ratio is larger for
small $T$.}
\end{figure}

Next, we show the CDF
really suppresses the transitions WFSS $\to$ NS and WFSS $\to$ SFSS by looking
at the projected atom numbers. For this purpose, we define the projection operators
$\Pi_{\rm W} = |{\rm WFSS}\rangle \langle {\rm WFSS}|,
\Pi_{\rm N} = |{\rm NS}\rangle \langle {\rm NS}|$ and
$\Pi_{\rm S} = |{\rm SFSS}\rangle \langle {\rm SFSS}|$ and evaluate the numbers
of atoms in these states as
$$
N_{\rm W} = \langle \Psi|\Pi_{\rm W}|\Psi \rangle,
N_{\rm N} = \langle \Psi|\Pi_{\rm N}|\Psi \rangle,
N_{\rm S} = \langle \Psi|\Pi_{\rm S}|\Psi \rangle.
$$
\begin{figure}
\begin{center}
\includegraphics[width=6cm]{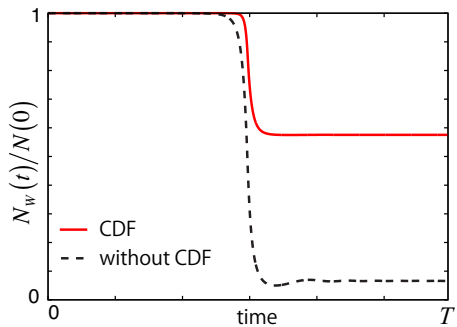}
\end{center}
\caption{(Color online) Instantaneous fraction $N_{\rm W} (t)/N(0)$ of atoms in WFSS
during vortex formation for $\log_{10}(T/\tau)=1.4$ and $r_0=2 a_{\rm HO}$.
The dashed and solid curves correspond to the controls without and with the CDF, respectively.}
\label{pop_com_stf0119}
\end{figure}
Figure 5 shows the time dependence of $N_W$ in the controls without CDF (dashed
curve) and with CDF (solid curve). The figure clearly shows that the
nonadiabatic transitions at $t \sim T/2$ are greatly suppressed and transitions
take place only at around $t=T/2$. This means that our approximation introduced in
 the denominator of (\ref{eq:bcd}) is not justifiable at $t \sim T/2$, where
 the original $\bs{B}_{\rm CD}$ diverges at $r=0$.
Otherwise, our simple approximation works reasonablly well.

Now an important observation is in order. In our original proposal, we need to prepare two sets
of Ioffe bars to produce the original confining quadrupole field and the counter-diabatic
magnetic field. Clearly this is demanding for experimentalists. All topological vortex formation
experiments so far were conducted with a single set of Ioffe bars.
Even if one could build a trap with two sets of Ioffe bars, aligning their centers exactly at the same place is practically impossible.
To circumvent this problem, we introduce
time-dependent gauge transformation so that the combined field $\bs{B}_{\perp}
+ \bs{B}_{\rm CD}$ is rotated
and the resulting field is in parallel to the original $\bs{B}_{\perp}$. Let
\begin{equation}
\alpha_B =\tan^{-1} \frac{|\tilde{\bs{B}}|}{|\bs{B}_{\perp}|}
 \end{equation}
be the angle between $\tilde{\bs{B}}$ and $\bs{B}_{\perp}$. If
 $\tilde{\bs{B}}$ is rotated by $-\alpha_B$, it becomes parallel to
$\bs{B}_{\perp}$ and hence it can be generated by a single set of Ioffe bars.
This rotation is implemented by the unitary transformation
\begin{equation}
U(\alpha_B) = e^{i \alpha_B F_z}.
\end{equation}
Now the Zeeman term of the Hamiltonian is transformed as
\begin{equation}
U(\alpha_B) \gamma \tilde{\bs{B}}\cdot \bs{F} U^{\dagger}(\alpha_B)
= \gamma \tilde{\bs{B}}_{\perp} \cdot \bs{F},
\end{equation}
where $\tilde{\bs{B}}_{\perp} = |\tilde{\bs{B}}| (x, -y,0)$. Of course, this is not the
whole story and there is a price we need to pay.
The Hamiltonian in the rotating frame acquires a gauge term
\begin{equation}
-i U \partial_t U^{\dagger} = - \dot{\alpha}_B F_z.
\end{equation}
This term, proportional to $F_z$, works as a magnetic field in the $z$-direction
and the bias field $B_z(t) = B_z(0)(1-2t/T)$ is replaced by
\begin{equation}
\tilde{B}_z(t) = B_z(0) \left(1-\frac{2t}{T}\right) - \dot{\alpha}_B.
\end{equation}
Note that $\alpha_B$ depends on time but not on space coordinates.
Numerical caluculation shows that the number of atoms in WFSS in the control with $\tilde{\bs{B}}_{\perp}$ and
$\tilde{B}_z$ is exactly the same as that for two sets of Ioffe bars for $t\ge T$. Figure 6
shows $\tilde{B}_z(t)$ and $|\tilde{\bs{B}}_{\perp}(t)| / |\bs{B}_{\perp}|$ for $^{23}$Na with
the parameters used in Fig. \ref{pop_com_stf0119}.
\begin{figure}
\begin{center}
\includegraphics[width=6cm]{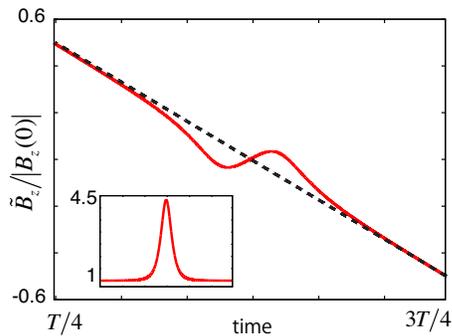}
\end{center}
\caption{(Color online)
Time dependence of $\tilde{B}_z(t)/B_z(0)$ (solid curve) and $B_z(t)$ (dashed line) for the same parameters used in Fig. \ref{pop_com_stf0119}.
The vertical and horizontal axes of the inset are $|\tilde{\bs{B}}_{\perp}(\bs{r},t)| / |\bs{B}_{\perp}(\bs{r})|$ and time with the range $T/4 \le t \le 3T/4$, respectively.}
\end{figure}

In summary, we have proposed a method to suppress nonadiabatic transitions while topological
vortex formation takes place in BEC of alkali atoms. The counter-diabatic field is generated
by a set of Ioffe bars, which is obtained by rotating the original Ioffe bars producing the
confining quadrupole field by $\pi/4$. Our numerical calculation demonstrates that
nondaiabatic transtions are suppressed for any inverstion time $T$ and,
in particular, it is most
impressive for a small $T$. We can further improve this scheme by applying a gauge
transformation to a rotating frame so that the combined field $\tilde{\bs{B}}$ is parallel to $\bs{B}_{\perp}$.
This requires modulation of $B_z(t)$ from linear time-dependence.
We believe our proposal is experimentally feasible by simple modifications of the
existing setup.

MN would like to thank Yuki Kawaguchi and Takeshi Kuwamoto for useful discussions.
The work of SM is partly supported by JSPS postdoc fellowship.
UG is supported in part by the NSF under Grant No.~Phy-1415600 and NSF-EPSCoR 1004094.
XC's work was partially supported by the NSFC (11474193 and 61176118),
the Shuguang and Pujiang Program (14SU35 and 13PJ1403000), the Specialized Research Fund for the Doctoral Program (2013310811003),
the Program for Eastern Scholar. MN's work is supported in part by a ``Topological
Quantum Phenomena'' Grant-in Aid for Scientific Research
on Innovative Areas (No. 22103003) from the Ministry of
Education, Culture, Sports, Science and Technology (MEXT)
of Japan. MN, TO and UG are also grateful to JSPS for partial support from
Grants-in-Aid for Scientific Research (Grant No. 26400422).

\end{document}